\documentclass[journal]{IEEEtran}
\usepackage{amsmath,amssymb,amsfonts,amsthm}
\usepackage{graphicx}
\usepackage{booktabs}
\usepackage{cite}
\usepackage{xcolor}
\usepackage{url}

\newtheorem{theorem}{Theorem}
\newtheorem{proposition}{Proposition}
\newtheorem{corollary}{Corollary}

\newcommand{\cX}{\mathcal{X}}
\newcommand{\cU}{\mathcal{U}}
\newcommand{\cY}{\mathcal{Y}}

\newcommand{\KL}{D_{\mathrm{KL}}}

\newcommand{\resultfigure}[1]{%
  \includegraphics[width=0.97\textwidth]{figures/#1}%
}

\title{Reliability Limits and Decoding for Partial Nanopore Protein Rereads With Persistent State}

\author{Hongbin~Ni,~\IEEEmembership{Graduate~Student~Member,~IEEE,}
        Haofan~Dong,~\IEEEmembership{Graduate~Student~Member,~IEEE,}
        and~Ozgur~B.~Akan,~\IEEEmembership{Fellow,~IEEE}%
\thanks{The authors are with the Centre for neXt Communications (CXC), Department of Engineering, University of Cambridge, Cambridge CB3~0FA, U.K.}%
\thanks{Ozgur~B.~Akan is also with the Centre for neXt Communications (CXC), Department of Electrical and Electronics Engineering, Ko\c{c} University, Istanbul 34450, T\"{u}rkiye.}}

\begin{document}
\bstctlcite{tcom_ref_control}
\maketitle

\begin{abstract}
Repeated observations of one physical object need not constitute independent channel uses. We model partial nanopore protein rereads as a finite-alphabet channel with canonical content, persistent readout, and pass-local coverage and synchronization. For exact compound-pass data, matched inference approaches the equivalence-class canonical posterior, and sitewise excess Bayes risk admits an action-aware achievable exponent. In an aligned specialization, observation-local redraw can cause linear-in-\(K\) growth in true-label negative log-likelihood (NLL). We derive order-\(b\) projection-stability bounds and an exact passwise-fusion diagnostic. On a PASTOR-informed semi-synthetic hard-symbol channel, label-blind deterministic-mixture importance sampling (LB--IS) agrees with exact enumeration at \(L=7\). At \(L=24,K=10\), LB--IS meets every prespecified aggregate absolute marginal-posterior and score-agreement criterion against a fixed high-allocation reference in three selected conditions. Joint agreement holds for the representative and high-NLL conditions, while the near-zero condition remains inconclusive. Exact \(L\le6\) benchmarks identify order 4 as the smallest tested common cap. At target scale, the reference supports selected unprojected functionals, while neither order 4 nor 5 attains joint agreement, defining a tested finite-memory boundary. Across 16 cells, the order-4 shared branch lowers NLL by \(0.033\)--\(0.224\) nats per residue relative to pass-local.
\end{abstract}

\begin{IEEEkeywords}
Molecular communication, nanopore protein sequencing, repeated observations, persistent-state channels, posterior reliability.
\end{IEEEkeywords}

\section{Introduction}
\IEEEPARstart{R}{epeated} measurements suppress local noise only under the correct dependence model. The assumption can fail when every measurement probes the same physical object. A molecule-specific state may persist while coverage, synchronization, and electrical noise vary from pass to pass. Multiplying posteriors formed under independent state redraw can then preserve a hard decision but distort its reported probability.

Repeated protein observations have been demonstrated at amino-acid resolution \cite{Brinkerhoff_2021}, and unfoldase-mediated slipping has yielded partial rereads \cite{Motone_2024} within the experimental constraints reviewed in \cite{Lu_2025}. PASTOR single-pass signals motivate the partial-reread setting \cite{Motone_2024,PastorData_2024}. An exploratory separability analysis selects the seven-residue subset \(\{\mathrm{G,Q,W,F,R,D,E}\}\) for the hard-symbol emission law, which serves as a controlled proxy for repeated-read inference. Because the records are aggregate and not molecule-linked, the upstream \(T/E\) allocation and repeated-pass persistence are semi-synthetic modeling choices.

Nanopore information models address signal memory, synchronization, and random response \cite{Mao_2018,McBain_2022}, while block-interference and block-fading models use states fixed over a block \cite{McEliece_1984,Biglieri_1998}. Repeated-read methods primarily target reconstruction, coding, template estimation, or storage rate \cite{Magner_2016,Chrisnata_2022,Lenz_2020,Srinivasavaradhan_2021,Wenger_2019,Deng_2024,Goyal_2026}. Molecular communication research addresses unknown-channel detection \cite{Jamali_2018,Qian_2021}, adaptive thresholds \cite{NiAkan_2025_ART_Rx}, cross-reactive receptors \cite{Civas_2024}, multiple absorbing receivers \cite{Ferrari_2022}, and diversity-based reaction-diffusion detection \cite{Lin_2023}. We instead study canonical-sequence inference through a nondegenerate molecule-persistent state under partial coverage and pass-local synchronization.

Posterior concentration under misspecification, Chernoff--Bhattacharyya discrimination, and mismatch analysis are classical \cite{Berk_1966,Kailath_1967,Merhav_1994}. We specialize these tools and the Kullback--Leibler (KL) chain rule to synchronized partial rereads. New elements are the compound-channel equivalence-class limit and excess-risk bound, projection stability after nonlinear multipass fusion, and diagnostics separating cross-pass from duplicate-local dependence. Unlike block-interference and block-fading models \cite{McEliece_1984,Biglieri_1998} and separate-read insertion--deletion--substitution (IDS) fusion \cite{Lenz_2020}, our model retains partial coverage and pass-local synchronization. Related DNA-storage work has a different state and inference target \cite{Goyal_2026}.

\begin{figure*}[!t]
\centering
\resultfigure{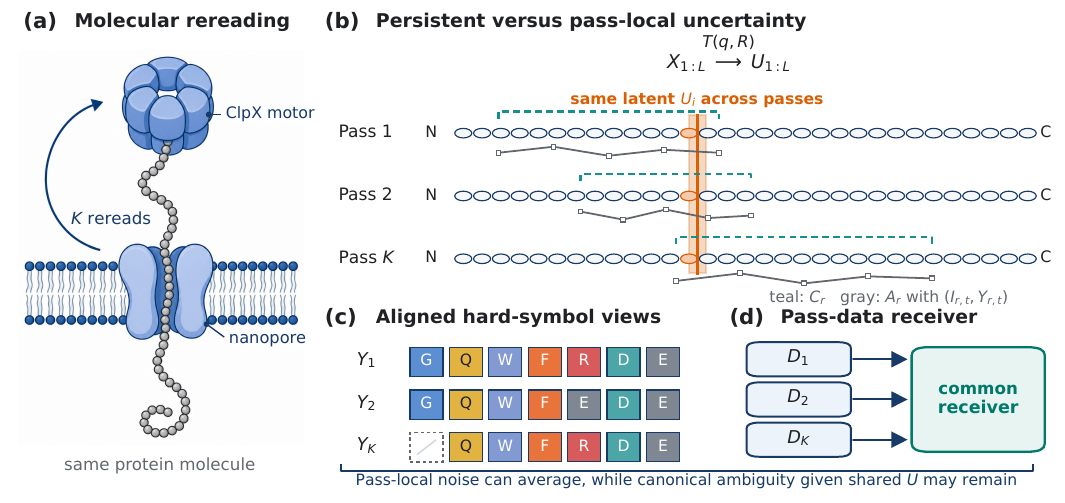}
\caption{Persistent-state partial-reread channel. (a) Conceptual ClpX--nanopore rereading over \(K\) passes. (b) \(X_{1:L}\) generates one persistent \(U_{1:L}\), while nuisance variables and \(D_r\) are pass-local. (c) Aligned hard-symbol views. The dashed \(Y_K\) cell is an alignment guide for an uncovered logical position, not an emitted erasure. (d) Common receiver.}
\label{fig:channel}
\end{figure*}

Fig.~\ref{fig:channel} summarizes a model that separates canonical content, a molecule-shared state, pass-local synchronization, and finite-alphabet outputs. Three nuisance-matched scopes define cross-pass and within-pass score contrasts. We derive an exact passwise-fusion diagnostic, canonical error limits, the observation-local true-label NLL slope despite possible MAP agreement, and a coverage-aware achievable excess-risk exponent. Exact compound-pass data yield matched posterior and Bayes-risk limits under partial coverage and synchronization.

Nonhomogeneous order-\(b\) projection provides tractable fused-odds and total-variation bounds. Exact enumeration and LB--IS assess selected unprojected functionals, while the \(L=24\) grid characterizes finite-order receivers, calibration, and emission sensitivity.

\section{Channel Model}
\label{sec:model}
\subsection{Persistent-state partial rereads}
Let \(\cX=\cU=\cY=\{\mathrm{G,Q,W,F,R,D,E}\}\) and \(X_{1:L}\in\cX^L\). Roman \(\mathrm R\) and \(\mathrm E\) are residue labels, whereas italic \(R\) and \(E\) denote probability matrices. The theory permits any positive prior \(p(x_{1:L})\). Numerically, sites are independent and identically distributed (i.i.d.) with \(p(x_{1:L})=\prod_i\pi_{x_i}\) and
\(\pi=(81,91,56,56,78,76,78)/516\) in \((\mathrm{G,Q,W,F,R,D,E})\) order. Conditional on \(X\), sites transform independently, \(p(u_{1:L}\mid x_{1:L})=\prod_iT_{x_i u_i}\), through
\begin{equation}
T_{xu}(q,R)=(1-q)\mathbf 1\{u=x\}+qR_{xu},
\label{eq:transition}
\end{equation}
where \(q\in[0,1]\), \(R_{xx}=0\), and \(R\) is row-stochastic. The scalar \(q\) is a persistent-deviation probability within a specified transition family, while the theory permits any fixed \(R\). Numerically, the grouped out-of-fold canonical-to-label law \(M^{\rm emp}\) is assigned to \(E\), and \(T(q,R)\) specifies the upstream persistent mechanism. Their product defines a semi-synthetic one-pass channel that is not fitted to \(M^{\rm emp}\). PASTOR identifies only the composite law, so partial-reread results are conditional on the \(T/E\) factorization. We test uniform \(R_{xu}=1/6\) and \(R_{xu}=E_x(u)/[1-E_x(x)]\) for \(u\ne x\). The weighted family is derived from \(E\), not independently measured. Here \(U\) is an effective molecule-persistent readout class, not an asserted biochemical mutation or directly observed state. The variable \(Y\) is the front end's hard label and \(E_u(y)\) its effective confusion law.

Pass \(r\) has coverage \(C_r\), observed endpoint pair \(e_r\), alignment path \(A_r\), visited indices \(I_{r,t}\), and symbols \(Y_{r,t}\). Write \(y_r=(Y_{r,t})_t\), set \(D_r=(e_r,y_r)\), and let bold lowercase letters collect pass-indexed families. For an inserted output, \(I_{r,t}=\bot\) and \(u_\bot\) is a dummy state with \(p(y\mid u_\bot,a)=|\cY|^{-1}\). Under a molecule-shared state \(U_{1:L}\), the general factorization is
\begin{align}
&p(x,u,\mathbf c,\mathbf e,\mathbf a,\mathbf i,\mathbf y)\nonumber\\
&=p(x)p(u\mid x)\prod_{r=1}^{K}\Bigg[
p(c_r)p(e_r\mid c_r)\nonumber\\
&\qquad{}\times p(a_r,i_r\mid c_r)
  \prod_t p(y_{r,t}\mid u_{i_{r,t}},a_{r,t})\Bigg].
\label{eq:factorization}
\end{align}
Marginalizing the pass-local coverage, endpoint, alignment, visited-index, and event variables in \eqref{eq:factorization} defines the exact compound-pass law
\begin{equation}
W_{\boldsymbol u}(d)=p(D_r=d\mid U_{1:L}=\boldsymbol u).
\label{eq:compound_pass_law}
\end{equation}
For a fixed channel setting, \(D_1,\ldots,D_K\) are conditionally i.i.d. given \(U_{1:L}\). The receiver assumes conditionally independent coverage and local nuisance draws across passes and marginalizes over bounded insertion, deletion, and duplication paths. Correlated pass coverage would require joint-pass inference rather than the product fusion used below. The aligned specialization additionally permits an observed null outcome whose probability depends on \(U\).

\begin{figure*}[!t]
\centering
\resultfigure{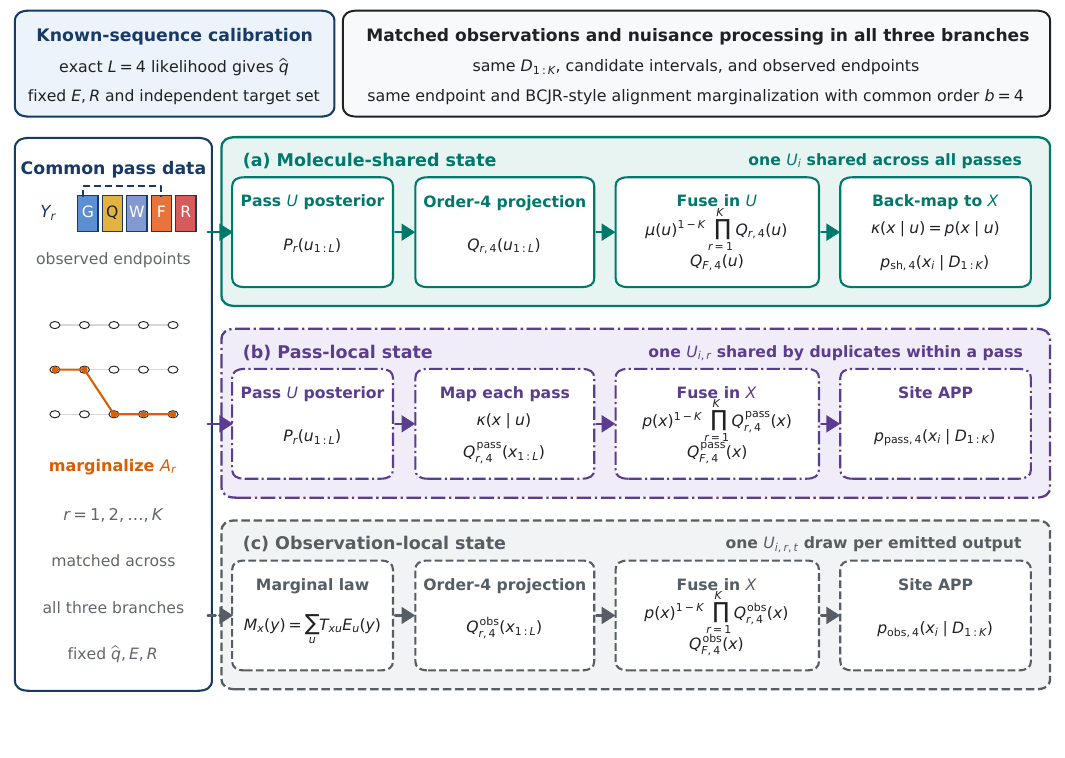}
\caption{Nuisance-matched receiver scopes. Known-sequence calibration molecules estimate \(\widehat q\) for fixed \(E,R\). Shared inference projects and fuses in \(U\) before the \(\kappa\) back-map, pass-local inference maps each pass to \(X\) before prior-corrected fusion, and observation-local inference uses \(M_x(y)\) directly. BCJR denotes Bahl--Cocke--Jelinek--Raviv. All branches share data, endpoint and alignment marginalization, and projection order. LB--IS is a numerical assessment method, not a receiver branch.}
\label{fig:receiver}
\end{figure*}

At the unprojected model level, Fig.~\ref{fig:receiver} summarizes three nuisance-matched receivers that share the molecule-shared law's observation set and differ only in persistence scope. They use identical endpoint support, alignment recursion, \(E\), \(R\), and \(q\), and are scored against the same realized \(X\).
\begin{align}
\text{molecule-shared:}\quad &U_i\sim T_{X_i}, \text{ once for all passes},\nonumber\\
\text{pass-local:}\quad &U_{i,r}\sim T_{X_i}, \text{ once within pass }r,\nonumber\\
\text{observation-local:}\quad &U_{i,r,t}\sim T_{X_i}, \text{ for each output}.
\label{eq:scopes}
\end{align}
The pass-local state is shared by duplicated outputs from one site in one pass and redrawn between passes. Thus
\begin{equation}
\underbrace{\mathrm{NLL}_{\mathrm{pass}}-\mathrm{NLL}_{\mathrm{shared}}}_{\text{cross-pass component}}
\quad\text{and}\quad
\underbrace{\mathrm{NLL}_{\mathrm{obs}}-\mathrm{NLL}_{\mathrm{pass}}}_{\text{duplicate-local component}}
\label{eq:contrasts}
\end{equation}
decompose the two dependence mechanisms at the unprojected model level. Their sum is the total observation-local versus molecule-shared contrast. At finite \(b\), the analogous score differences are receiver-level diagnostics that also contain branch-dependent projection error.

\subsection{Aligned specialization}
For one covered site, let \(E_u(a)=p(Y=a\mid U=u)\), and let \(N(a)\) count occurrences of \(a\) among \(n\) aligned outputs. The molecule-shared posterior has unnormalized weight
\begin{equation}
W_n(x)=\pi_x\sum_u T_{xu}\prod_{a\in\cY}E_u(a)^{N(a)}.
\label{eq:shared_weight}
\end{equation}
The observation-local model uses
\begin{equation}
M_x(a)=\sum_uT_{xu}E_u(a)
\label{eq:marginal_emission}
\end{equation}
for each output. With one output per pass, the pass-local and observation-local laws coincide. Duplications distinguish them.

For \(K\) aligned outputs \(y_{1:K}\), the shared-state and locally redrawn likelihoods are
\begin{align}
L_{\rm sh}(x,y_{1:K})&=\sum_uT_{xu}\prod_{k=1}^{K}E_u(y_k),\nonumber\\
L_{\rm loc}(x,y_{1:K})&=\prod_{k=1}^{K}M_x(y_k).
\label{eq:shared_local_likelihoods}
\end{align}
Prior-corrected multiplication of canonical one-pass posteriors produces \(L_{\rm loc}\), not \(L_{\rm sh}\). The resulting canonical posteriors coincide if and only if their likelihood vectors have the same support over positive-prior labels and \(L_{\rm sh}(x,y_{1:K})/L_{\rm loc}(x,y_{1:K})\) is constant on that support. This follows by equating their normalized weights \(\pi_xL_{\rm sh}(x,y_{1:K})\) and \(\pi_xL_{\rm loc}(x,y_{1:K})\). Equality is guaranteed for \(K=1\), when \(U\) is deterministic given \(X\), or when all emission rows on each support of \(T_x\) coincide, but not in general.

The one-pass law \(M=TE\) admits multiple unrestricted \(T/E\) factorizations.

For fixed \(\bar M=M^{\rm emp}\), \((I,\bar M)\) and \((\bar M,I)\) have different two-pass laws despite the same one-pass law. More generally, \(T_\alpha=(1-\alpha)I+\alpha\bar M\) and \(E_\alpha=T_\alpha^{-1}\bar M\) are stochastic factorizations for \(\alpha=0,0.1,0.25,1\). Exact count-vector enumeration shows increasing NLL gap and mean posterior TV with \(K\) for every tested \(\alpha>0\). At \(K=10\), the gaps are \(0,0.276,0.828,6.049\) nats per site and TV values are \(0,0.036,0.091,0.375\). Molecule-linked rereads provide information about the allocation, while unrestricted identification requires structural assumptions. The present study evaluates fixed \(E=M^{\rm emp}\) and the two specified \(R\) families.

\section{Reliability Analysis and Receiver Design}
\label{sec:analysis}
\subsection{Asymptotic reliability}
Write \(J_{xu}=\pi_xT_{xu}\), \(p_U(u)=\sum_xJ_{xu}\), and \(\cU_+=\{u:p_U(u)>0\}\). Let \(B_k\) denote the observed coverage indicator in pass \(k\), and let \(\mathcal D_K=(Y_{1:N_K},B_{1:K})\) collect the covered outputs and indicators through pass \(K\). Define the finite-pass canonical Bayes error \(P_{e,K}=1-\mathbb E[\max_{x\in\cX}p(X=x\mid\mathcal D_K)]\). All logarithms are natural. Unless stated otherwise, probabilities and expectations in this section are under the true joint channel law.

\begin{theorem}[Shared-state posterior and limits]
\label{thm:shared}
Assume finite alphabets and pairwise distinct emission rows \(E_u\) for \(u\in\cU_+\). Conditional on \(U=u\), covered outputs are i.i.d. with law \(E_u\). Coverage indicators are observed, independent of \((X,U)\), and their cumulative count \(N_K\) diverges almost surely. Under the molecule-shared model, conditional on \(U=u\in\cU_+\),
\begin{equation}
p(X=x\mid \mathcal D_K)\longrightarrow\frac{J_{xu}}{p_U(u)}
\end{equation}
almost surely. The canonical Bayes error and expected NLL satisfy
\begin{align}
\lim_{K\to\infty}P_{e,K}=P_{e,\infty}
&=1-\sum_{u\in\cU_+}\max_{x\in\cX}J_{xu},
\label{eq:error_floor}\\
\lim_{K\to\infty}\mathbb E[-\log p(X\mid \mathcal D_K)]
&=H(X\mid U).
\label{eq:shared_nll_limit}
\end{align}
\end{theorem}

Once \(U\) is identified, the remaining canonical uncertainty is \(H(X\mid U)\). It is positive exactly when \(p(X\mid U)\) is nondegenerate, equivalently when no function of \(U\) recovers \(X\) almost surely. At \(q=0\), \(U=X\) and the floor vanishes. Coincident emission rows instead identify a single-site equivalence class \(V_{\rm a}\), giving limits \(p(X\mid V_{\rm a})\), \(H(X\mid V_{\rm a})\), and its aggregated error floor.

For the observation-local decoder, equivalently encode \(\mathcal D_K\) by \(O_{1:K}\), where \(O_k=\varnothing\) when uncovered and \(O_k=Y_k\) otherwise. Using KL divergence, define \(d_x(u)=\KL(E_u\Vert M_x)\), \(d_\star(u)=\min_zd_z(u)\), and \(\mathcal S(u)=\{x:d_x(u)=d_\star(u)\}\).

\begin{theorem}[Observation-local pseudo-truth and NLL slope]
\label{thm:local}
Suppose the observed coverage indicators are i.i.d. Bernoulli-\(h\), \(h>0\), independent of \((X,U)\). Conditional on \(U=u\), covered outputs are i.i.d. with law \(E_u\) and independent of coverage. Let the decoder use the matching observed-coverage law, a positive prior, and \(M_x\) for every covered output. Assume \(E_u\ll M_x\) whenever \(J_{xu}>0\). Conditional on \(U=u\), the posterior mass assigned to \(\mathcal S(u)\) tends to one almost surely. Define
\begin{equation}
\Gamma=\sum_{\{(x,u):J_{xu}>0\}}J_{xu}[d_x(u)-d_\star(u)].
\label{eq:gamma}
\end{equation}
Then
\begin{equation}
\mathbb E[-\log p_{\mathrm{obs}}(X\mid O_{1:K})]
=Kh\Gamma+o(K).
\label{eq:local_nll_slope}
\end{equation}
\end{theorem}

The true-label slope permits \(\Gamma=0\), which holds exactly when \(x\in\mathcal S(u)\) for every \(J_{xu}>0\), although finite-\(K\) posteriors may still differ. A nonunique minimizer yields set, not point, concentration. If the theorem's support condition fails, expected NLL is infinite. With one output per pass, pass-local and observation-local likelihoods coincide. Even when observation-local and molecule-shared MAP actions agree, the observation-local decoder may drive a still-possible true label exponentially toward zero, so accuracy cannot reveal the mismatch.

For state-dependent coverage, augment one-pass emissions by
\begin{equation}
\widetilde E_u(\varnothing)=1-h_u,\qquad
\widetilde E_u(a)=h_uE_u(a).
\label{eq:augmented}
\end{equation}
Let \(a(u)=\arg\max_xJ_{xu}\) be unique and define
\begin{align}
C_{uv}&=-\log\inf_{0\le t\le1}
\sum_o\widetilde E_u(o)^{1-t}\widetilde E_v(o)^t,\nonumber\\
C_{\mathrm{dec}}&=\min_{\substack{u<v,\ u,v\in\cU_+\\a(u)\ne a(v)}}C_{uv}.
\label{eq:cdec}
\end{align}
The symbol \(<\) denotes any fixed ordering of the finite state alphabet. The minimum over an empty set is \(+\infty\).

\begin{theorem}[Coverage-aware achievable exponent]
\label{thm:coverage}
Suppose augmented pass outcomes are conditionally i.i.d. given \(U\), and \(a(u)=\arg\max_xJ_{xu}\) is unique for every \(u\in\cU_+\). Let \(\widehat U_K\) be a MAP state estimate based on \(O_{1:K}\), and set \(\mathcal R_K^{\rm 2st}=\Pr\{a(\widehat U_K)\ne X\}\), \(\mathcal R_K^{\rm Bayes}=P_{e,K}\), and \(\mathcal R_\infty=P_{e,\infty}\), the oracle error with \(U\) observed. Then
\begin{equation}
\liminf_{K\to\infty}-\frac1K
\log(\mathcal R_K^{\rm 2st}-\mathcal R_\infty)\ge C_{\mathrm{dec}},
\label{eq:achievable_exponent}
\end{equation}
with exponent \(+\infty\) when the excess is eventually zero. The direct canonical Bayes risk obeys \(0\le\mathcal R_K^{\rm Bayes}-\mathcal R_\infty\le\mathcal R_K^{\rm 2st}-\mathcal R_\infty\), and therefore achieves at least this exponent.
\end{theorem}

Only state pairs inducing different canonical actions enter the achievable exponent \(C_{\mathrm{dec}}\). No converse is established. Applying the classical Bhattacharyya choice \(t=1/2\) \cite{Kailath_1967} to common coverage gives the partial-coverage bound. State-dependent null probabilities contribute evidence through \(\widetilde E_u(\varnothing)\), so discarding observed nulls can be suboptimal when \(h_u\) varies with state.

\begin{proposition}[Compound-pass reliability]
\label{prop:compound}
Assume finite sequence and pass-data alphabets, \(X_{1:L}\to U_{1:L}\to D_{1:\infty}\), and conditionally i.i.d. \(D_1,D_2,\ldots\) given \(U_{1:L}=\boldsymbol u\), with law \(W_{\boldsymbol u}\). Let \(\cU_L^+=\{\boldsymbol u:p(U_{1:L}=\boldsymbol u)>0\}\). On \(\cU_L^+\), define \(\boldsymbol u\sim\boldsymbol v\) when \(W_{\boldsymbol u}=W_{\boldsymbol v}\), let \(V=[U_{1:L}]\) be the resulting pass-observational equivalence class with alphabet \(\mathcal V\), and write \(W_c\) for the common law in class \(c\). Conditional on any \(c\in\mathcal V\) with \(p(V=c)>0\), the exact molecule-shared posterior satisfies
\begin{equation}
p(X_{1:L}=\boldsymbol x\mid D_{1:K})
\longrightarrow p(X_{1:L}=\boldsymbol x\mid V=c)
\label{eq:compound_posterior}
\end{equation}
almost surely. The limiting sitewise Bayes error and marginal NLL per residue are
\begin{equation}
\mathcal R_{\infty}^{\rm site}
=1-\frac1L\sum_{i=1}^{L}\sum_{c\in\mathcal V}
\max_{a\in\cX}p(X_i=a,V=c).
\label{eq:compound_error}
\end{equation}
\begin{multline}
\lim_{K\to\infty}\frac1L\sum_{i=1}^{L}
\mathbb E[-\log p(X_i\mid D_{1:K})]\\
=\frac1L\sum_{i=1}^{L}H(X_i\mid V).
\label{eq:compound_nll}
\end{multline}
If the \(W_{\boldsymbol u}\) are pairwise distinct, then \(V=U_{1:L}\). Under the i.i.d. site prior and memoryless \(X_i\to U_i\) law used numerically, \eqref{eq:compound_nll} then reduces to \(H(X\mid U)\) per residue.
\end{proposition}

\begin{corollary}[Compound-pass achievable excess-risk exponent]
\label{cor:compound_exponent}
Under Proposition~\ref{prop:compound}, suppose the sitewise Bayes action
\(a_i(c)=\arg\max_{a\in\cX}p(X_i=a\mid V=c)\) is unique for all \(i,c\). Set \(a(c)=(a_1(c),\ldots,a_L(c))\) and define
\begin{align}
C^W_{cc'}&=-\log\inf_{0\le t\le1}
\sum_d W_c(d)^{1-t}W_{c'}(d)^t,\nonumber\\
C^W_{\rm dec}&=
\min_{\substack{c,c'\in\mathcal V,\ c<c'\\a(c)\ne a(c')}}C^W_{cc'},
\label{eq:compound_dec_exponent}
\end{align}
where \(<\) denotes the same fixed ordering convention and an empty minimum is \(+\infty\). Let \(\widehat V_K\) be the MAP class estimate. Define \(\mathcal R_{K,W}^{\rm 2st}=L^{-1}\sum_i\Pr\{a_i(\widehat V_K)\ne X_i\}\) and \(\mathcal R_{K,W}^{\rm Bayes}=1-L^{-1}\sum_i\mathbb E[\max_{a\in\cX}p(X_i=a\mid D_{1:K})]\). Then
\begin{equation}
\liminf_{K\to\infty}-\frac1K\log
\left(\mathcal R_{K,W}^{\rm 2st}
-\mathcal R_\infty^{\rm site}\right)
\ge C^W_{\rm dec},
\label{eq:compound_achievable}
\end{equation}
with exponent \(+\infty\) when the excess is eventually zero. The direct sitewise Bayes receiver obeys
\(0\le\mathcal R_{K,W}^{\rm Bayes}-\mathcal R_\infty^{\rm site}
\le\mathcal R_{K,W}^{\rm 2st}-\mathcal R_\infty^{\rm site}\), and therefore achieves no smaller excess-risk exponent.
\end{corollary}

For the target mechanism below, distinct \(W_{\boldsymbol u}\) give \(V=U_{1:L}\). Table~\ref{tab:scope} summarizes the analytical scopes, and the nested-\(K\), \(L=4\) study illustrates finite-pass convergence without estimating \(C^W_{\rm dec}\).

\begin{table*}[!t]
\caption{Analytical objects and nuisance treatment}
\label{tab:scope}
\centering
\footnotesize
\setlength{\tabcolsep}{3.5pt}
\renewcommand{\arraystretch}{1.00}
\begin{tabular}{p{0.17\textwidth}p{0.26\textwidth}p{0.47\textwidth}}
\toprule
Analysis & Exact object & Numerical link and nuisance treatment \\
\midrule
Theorems~\ref{thm:shared}--\ref{thm:coverage}, Fig.~\ref{fig:theory} & Aligned single-site outcomes & Isolates the posterior limit, NLL slope, and coverage effect in the aligned model \\
Prop.~\ref{prop:compound}, Cor.~\ref{cor:compound_exponent}, pass-law distinctness check & Exact finite pass data \(D_r=(e_r,y_r)\) & Marginalizes intervals, paths, and local events. Nested \(K\) illustrates finite-pass convergence \\
Theorem~\ref{thm:order}, exact and target-scale checks & Arbitrary finite exact pass posterior \(P_r\) & Enumeration checks order projection through \(L=7\). At \(L=24\), LB--IS and the high-allocation reference assess selected unprojected functionals and finite-order deviation from that reference \\
Three-scope study & Length-\(24\) partial sequence with fitted \(\widehat q\) and order \(b\) & Random intervals, noisy endpoints, bounded path marginalization, and explicit local events \\
\bottomrule
\end{tabular}
\end{table*}

Coincident emission rows identify only an equivalence class. Support mismatch can make observation-local NLL or \(\Omega_b\) infinite, and state-dependent nulls must be retained. Calibration fixes \(E,R\) and transfers \(q\) across lengths. Numerical scores use an i.i.d. canonical prior, while the theory permits any positive prior.

\begin{figure*}[!t]
\centering
\resultfigure{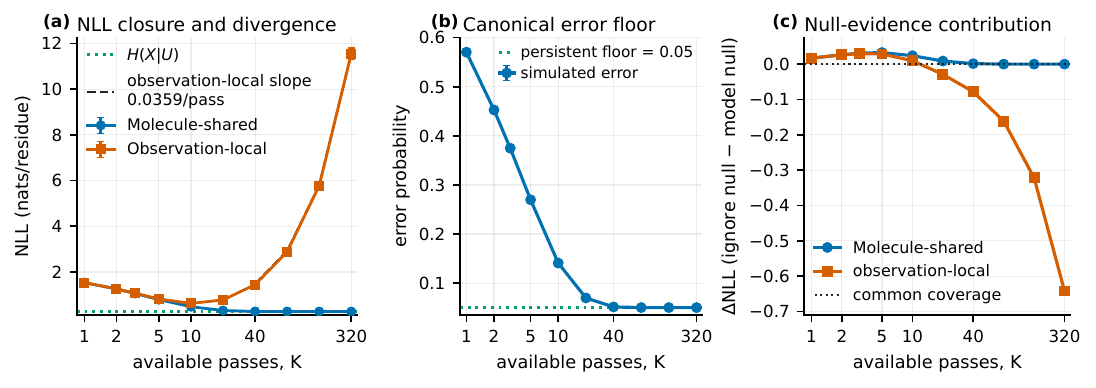}
\caption{Aligned reliability at \(q=0.05\), confusability-weighted \(R\), and \(200\,000\) sites. (a) At \(h=0.60\), shared NLL approaches \(H(X\mid U)=0.272\), while observation-local NLL has slope \(h\Gamma=0.0359\) nats per pass, with \(0.47\%\) maximum relative discrepancy. (b) Error approaches the ambiguity floor. (c) State-dependent observed nulls alter evidence, whereas common state-independent coverage cancels.}
\label{fig:theory}
\end{figure*}

\subsection{Order-\(b\) projection and the three receivers}
For each pass, \(D_r=(e_r,y_r)\) contains the observed endpoints and hard-symbol sequence. A semi-Markov extension of the Bahl--Cocke--Jelinek--Raviv (BCJR) forward--backward recursion \cite{Bahl_1974,McBain_2022} gives the exact molecule-shared pass posterior \(P_r(u_{1:L})=p(u_{1:L}\mid D_r)\). Its nonhomogeneous order-\(b\) projection is
\begin{equation}
Q_{r,b}(u_{1:L})=P_r(u_{1:b})
\prod_{i=b+1}^{L}P_r(u_i\mid u_{i-b:i-1}).
\label{eq:order_b}
\end{equation}
For \(b=0\), the prefix is empty. A conditional distribution following a zero-probability context may be chosen arbitrarily because that context contributes zero to forward KL. For \(b=L-1\), the representation is exact. With the positive sequence prior \(\mu(\boldsymbol u)=\sum_{\boldsymbol x\in\cX^L}p(\boldsymbol x)p(\boldsymbol u\mid\boldsymbol x)\), the exact and projected molecule-shared fusions are
\begin{align}
P_F(\boldsymbol u)&\propto\mu(\boldsymbol u)^{1-K}\prod_{r=1}^{K}P_r(\boldsymbol u),\nonumber\\
Q_{F,b}(\boldsymbol u)&\propto\mu(\boldsymbol u)^{1-K}\prod_{r=1}^{K}Q_{r,b}(\boldsymbol u).
\label{eq:fusion}
\end{align}
The projected molecule-shared canonical a posteriori probability (APP) is
\begin{align}
p_{\mathrm{sh},b}(\boldsymbol x\mid D_{1:K})
&=\sum_{\boldsymbol u}\frac{p(\boldsymbol x)p(\boldsymbol u\mid\boldsymbol x)}{\mu(\boldsymbol u)}Q_{F,b}(\boldsymbol u).
\label{eq:shared_app}
\end{align}

Under the i.i.d. canonical prior used numerically, the pass-local receiver lets \(\lambda_r(u)=p(y_r\mid e_r,u)\) after interval and alignment marginalization, then defines
\begin{align}
P_r^{\mathrm{pass}}(x)
&=\sum_u P_r(u)\prod_{i=1}^{L}\kappa(x_i\mid u_i),\nonumber\\
\kappa(x_i\mid u_i)
&=\frac{\pi_{x_i}T_{x_i u_i}}{p_U(u_i)}.
\label{eq:pass_posterior}
\end{align}
The receiver maps each contiguous clique by
\begin{equation}
Q_{r,b}^{\mathrm{pass}}(x_{i-b:i})
=\sum_{u_{i-b:i}}P_r(u_{i-b:i})
\prod_{j=i-b}^{i}\kappa(x_j\mid u_j),
\label{eq:clique_backmap}
\end{equation}
then reconstructs the order-\(b\) canonical chain. This is the order-\(b\) projection of \eqref{eq:pass_posterior} because \(Q_{r,b}\) preserves every required \((b+1)\)-site marginal of \(P_r\). For the observation-local receiver, construct its canonical pass posterior with the same recursion after replacing each state-conditioned output factor by \(M_{x_i}(y)\). Project the pass-local and observation-local canonical posteriors to \(Q_{r,b}^{s}\), \(s\in\{\mathrm{pass},\mathrm{obs}\}\), and fuse
\begin{equation}
Q_{F,b}^{s}(x)\propto p(x)^{1-K}\prod_{r=1}^{K}Q_{r,b}^{s}(x).
\label{eq:canonical_fusion}
\end{equation}
The unprojected receiver definitions change only persistence scope. At finite \(b\), however, the shared branch projects in \(U\)-space before the \(\kappa\) back-map, whereas the pass-local and observation-local branches project in \(X\)-space. Equal nominal order therefore does not equalize branchwise approximation error because projection, back-mapping, and nonlinear fusion need not commute. Under the pass-local model, duplicated emissions are independent conditional on their common \(U_{i,r}\) but dependent after marginalization. Under the observation-local model, the two state draws are independent and each output contributes its own \(M_x\) factor.

\begin{theorem}[Order-\(b\) projection and fusion stability]
\label{thm:order}
For \(0\le b\le L-1\) and any pass posterior \(P_r\) on a finite sequence alphabet, \(Q_{r,b}\) is a forward-KL minimizer over nonhomogeneous order-\(b\) chains and
\begin{equation}
\KL(P_r\Vert Q_{r,b})
=\sum_{i=b+1}^{L}
\operatorname{CMI}_{P_r}(U_i,U_{1:i-b-1}\mid U_{i-b:i-1}).
\label{eq:cmi_residual}
\end{equation}
Here \(\operatorname{CMI}_P(A,B\mid C)\) denotes conditional mutual information under \(P\).
The residual is nonincreasing in \(b\) and vanishes at \(b=L-1\). If \(P_r\) and \(Q_{r,b}\) are strictly positive, let \(\operatorname{osc}f=\max_uf(u)-\min_uf(u)\) and
\begin{equation}
\Omega_b=\sum_{r=1}^{K}\operatorname{osc}
\log\frac{P_r}{Q_{r,b}}.
\label{eq:omega}
\end{equation}
For any \(u,v\),
\begin{align}
\left|\log\frac{P_F(u)}{P_F(v)}
-\log\frac{Q_{F,b}(u)}{Q_{F,b}(v)}\right|&\le\Omega_b,
\label{eq:odds_bound}\\
\operatorname{TV}(P_F,Q_{F,b})&\le\tanh(\Omega_b/4).
\label{eq:tv_bound}
\end{align}
\end{theorem}

The shared \(U\!\to X\) back-map contracts TV, while Theorem~\ref{thm:order} applies directly to the other canonical posteriors. Its KL residual is an average quantity and \(\Omega_b\) is worst case, so small residuals or larger \(b\) need not yield a tight TV bound. At \(L=4,K=10,b=2\), the bound is \(0.796\), versus canonical-sequence TV \(0.0162\), illustrating its conservatism. The bound therefore does not certify the \(L=24\) orders. Exact short-chain APP, regret, and contrast errors guide order selection, and target-scale agreement is assessed numerically. Given the passwise moments and prior order at most \(b\), fusion and marginalization cost \(O(KL|\cU|^{b+1})\). Alignment has a separate pass-likelihood cost.

\subsection{Fixed-family known-sequence calibration and identifiability}
For an independent calibration molecule \(s\), let \(x^{(s)}_{1:4}\) be its known sequence and \(D^{(s)}_{1:K}\) its \(K\) passes. Conditional on the observed endpoint pairs \(\boldsymbol e^{(s)}\), independence across passes gives
\begin{equation}
p(\boldsymbol y^{(s)}\mid \boldsymbol e^{(s)},u_{1:4})
=\prod_{r=1}^{K}\lambda_{s,r}(u_{1:4}),
\label{eq:reference_pass_product}
\end{equation}
where \(\lambda_{s,r}=p(y^{(s)}_r\mid e^{(s)}_r,u_{1:4})\) marginalizes the candidate interval and local alignment using the endpoint-conditioned interval law. Because endpoint noise is independent of \(q\), \(X\), and \(U\) in the specified assay, its likelihood factor is constant in \(q\) and can be conditioned out. The exact conditional object likelihood is
\begin{equation}
\mathcal L_s(q)=
\sum_{u_{1:4}}
p(u_{1:4}\mid x^{(s)}_{1:4},q,R)
p(\boldsymbol y^{(s)}\mid \boldsymbol e^{(s)},u_{1:4}).
\label{eq:reference_likelihood}
\end{equation}
The second factor therefore uses the length-four analog of the target receiver's endpoint-conditioned interval kernel, insertion, deletion, duplication, alignment, and emission mechanisms. We estimate
\begin{equation}
\widehat q=\arg\max_{q\in[0,0.20]}
\sum_s\log\mathcal L_s(q).
\label{eq:qhat}
\end{equation}
We maximize by enumerating all \(7^4\) latent sequences, evaluating 4001 points on \([0,0.20]\), and applying bounded refinement without assuming concavity. The calibration assay uses moderate events, mild endpoint noise, and \(K=10\). Its 200 length-four sequences form a randomly permuted 800-symbol multiset with counts differing by at most one. Span probabilities for lengths 1--4 are \((0.20,0.35,0.15,0.30)\), with valid starts uniform conditional on span. Each \((q,R,\text{replicate})\) estimate is reused across four target nuisance cells. Transfer to \(L=24\) holds \(E\), \(R\), and the \(q\) family fixed, with calibration estimating \(q\) within that family.

An aligned two-output specialization gives a compact identifiability check for that fixed family. Let \(G_q(y,y')\) be the joint law of two covered outputs that share one effective state. Under the i.i.d. canonical prior,
\begin{align}
G_q&=G_0+q\Delta_R,\label{eq:pair_affine}\\
G_0(y,y')&=\sum_x\pi_xE_x(y)E_x(y'),\nonumber\\
\Delta_R(y,y')&=\sum_x\pi_x
\left[\sum_uR_{xu}E_u(y)E_u(y')-E_x(y)E_x(y')\right].
\nonumber
\end{align}

\begin{proposition}[Fixed-family pair identifiability]
\label{prop:pair}
Assume \((\pi,E,R)\) are known, fixed, and common to the included aligned calibration pairs. Each included ordered pair shares one \(U\), its outputs are conditionally independent given \(U\), and its inclusion is independent of \((X,U)\), so its law is \(G_q\). The scalar \(q\) is identifiable from \(G_q\) if and only if \(\|\Delta_R\|_F>0\). If \(\widehat G\) is the empirical ordered-pair distribution, then
\begin{equation}
\widetilde q=
\frac{\langle\widehat G-G_0,\Delta_R\rangle}
{\|\Delta_R\|_F^2}
\label{eq:pair_estimator}
\end{equation}
is unbiased before truncation to the admissible \(q\) interval.
\end{proposition}

This diagnostic establishes identifiability of \(q\) within the fixed family and does not replace the synchronization-aware object likelihood \eqref{eq:reference_likelihood}.

\section{Numerical Evaluation}
\label{sec:numerical}
Numerical evaluation validates LB--IS against exact \(L=7\) enumeration and then assesses selected unprojected functionals at fixed \(L=24\) conditions against a high-allocation reference. The complementary 16-cell grid characterizes calibrated finite-order receiver behavior and emission sensitivity, not exact persistence-scope effects.

\subsection{Numerical design}
The seven-class \(E\) uses grouped out-of-fold predictions from 516 PASTOR records \cite{Motone_2024,PastorData_2024}. Using five-fold stratification that keeps 215 acquisition groups intact, we fit a balanced 150-tree random forest with 14 current and shape features, square-root feature subsampling, and minimum leaf size two. Grouped accuracy and macro recall are \(67.05\%\) and \(67.01\%\), while per-class recall ranges from \(48.2\%\) to \(81.6\%\). Whole-run grouping lowers accuracy to \(61.05\%\). Because alphabet selection and evaluation use the same collection, these values are conditional. Independent data with the alphabet fixed, or nested outer-fold selection, would provide a selection-unbiased assessment. The emission law \(E\) uses add-one-smoothed confusion counts, \(E_u(y)=(N_{uy}+1)/(n_u+7)\) for \(n_u=\sum_yN_{uy}\), as in Table~\ref{tab:emission}.

\begin{table*}[!t]
\caption{Add-one-smoothed PASTOR-informed emission laws \(E_u(y)\). The acquisition-grouped law is primary. The whole-run-blocked law provides the crossed-law sensitivity comparison.}
\label{tab:emission}
\centering
\footnotesize
\setlength{\tabcolsep}{2.25pt}
\begin{tabular}{c*{7}{r}@{\hspace{1.8em}}c*{7}{r}}
\toprule
\multicolumn{8}{c}{Acquisition-grouped out-of-fold law} &
\multicolumn{8}{c}{Whole-run-blocked law} \\
\cmidrule(r){1-8}\cmidrule(l){9-16}
\(u\backslash y\) & G & Q & W & F & R & D & E &
\(u\backslash y\) & G & Q & W & F & R & D & E \\
\midrule
G & .614 & .170 & .057 & .011 & .034 & .023 & .091 &
G & .580 & .170 & .057 & .023 & .034 & .011 & .125 \\
Q & .133 & .531 & .071 & .071 & .163 & .010 & .020 &
Q & .122 & .480 & .082 & .122 & .153 & .010 & .031 \\
W & .048 & .095 & .698 & .032 & .079 & .016 & .032 &
W & .079 & .063 & .683 & .032 & .095 & .032 & .016 \\
F & .063 & .159 & .175 & .444 & .127 & .016 & .016 &
F & .079 & .159 & .159 & .460 & .111 & .016 & .016 \\
R & .012 & .082 & .059 & .047 & .741 & .012 & .047 &
R & .012 & .118 & .082 & .118 & .612 & .012 & .047 \\
D & .036 & .012 & .012 & .012 & .012 & .759 & .157 &
D & .024 & .012 & .012 & .012 & .012 & .687 & .241 \\
E & .094 & .106 & .012 & .012 & .024 & .176 & .576 &
E & .106 & .071 & .012 & .035 & .012 & .259 & .506 \\
\bottomrule
\end{tabular}
\end{table*}

At \(L=4,5,6\), the 16-cell factorial studies use 100, 50, and 20 molecules per cell, respectively. They use \(K\in\{1,2,5,10\}\), \(b\in\{0,1,2,3,L-1\}\), and \(b=4\) at \(L=6\), with exact posterior enumeration. A target span \(s\) maps to \(\widetilde s=\max\{1,\min[L,\lfloor sL/24+1/2\rfloor]\}\), with coincident masses combined and valid starts uniform. The \(L=7\) order assessment uses two molecules per cell, weighted \(R\), moderate endpoints, both \(q\) values and event regimes, \(K\in\{5,10\}\), and \(b\in\{3,4,6\}\). A fixed benchmark compares LB--IS with exact \(L=7\) enumeration. At \(L=24\), LB--IS and paired \(b=4,5\) receivers are compared with the high-allocation unprojected-model reference.

\subsection{Synchronization likelihood and score construction}
For observed endpoint pair \(e_r\), candidate interval \(c_r\), alignment path \(a_r\), and visited-index sequence \(i_r\), the state-conditioned pass likelihood is
\begin{align}
\lambda_r(u_{1:L})
&=\sum_{c_r}p(c_r\mid e_r)
\sum_{a_r,i_r}p(a_r,i_r\mid c_r)\nonumber\\
&\quad\times\prod_t
p(y_{r,t}\mid u_{i_{r,t}},a_{r,t}).
\label{eq:pass_likelihood}
\end{align}
At each covered site, a Bernoulli-\(p_{\rm ins}\) insertion draw precedes deletion, ordinary emission, or duplication with probabilities \(p_{\rm del}\), \(1-p_{\rm del}-p_{\rm dup}\), and \(p_{\rm dup}\). The recursion sums all supported paths. Insertions are uniform. Ordinary and duplicate events draw once or twice independently from the state row of \(E\). Finite interval and endpoint supports with bounded events give \(|y_r|\le3L\). Since \(\cY\) is finite, \(D_r=(e_r,y_r)\) has the finite alphabet required by Proposition~\ref{prop:compound}.

The target pass laws are distinguishable. Let \(\mathcal A=\{|Y_r|=3L\}\). Full-span coverage has positive mass and \(p_{\rm ins},p_{\rm dup}>0\), so \(\mathcal A\) has positive state-independent probability. Because each covered site emits at most three symbols, \(\mathcal A\) forces full coverage, one insertion, and one duplication per site. Conditional on \(\mathcal A\), the ordered symbol law is
\begin{equation}
\bigotimes_{i=1}^{L}
\left[\operatorname{Unif}(\cY)\otimes E_{u_i}\otimes E_{u_i}\right].
\label{eq:distinguishing_event}
\end{equation}
For \(\boldsymbol u\ne\boldsymbol v\), a differing coordinate exposes distinct rows of Table~\ref{tab:emission}. Thus \(W_{\boldsymbol u}\ne W_{\boldsymbol v}\) and \(V=U_{1:L}\), establishing target-channel asymptotic identifiability. Finite-\(K\) behavior is examined below.

For 100 representative \(L=4\) molecules with known \(q=0.10\), exact shared NLL falls from \(1.706\) at \(K=1\), through \(1.472\) and \(1.048\) at \(K=2,5\), to \(0.771\) at \(K=10\). The paired reduction is \(0.935\) with 95\% interval \([0.815,1.055]\). At \(K=10\), NLL remains above \(H(X\mid U)=0.473\), and exact pass-local minus shared NLL is \(0.177\) \([0.083,0.271]\). These results demonstrate finite-pass convergence and a positive exact contrast.

The interval prior draws a span component, an integer span within it, and a valid start. With endpoint-offset mass \(g\), the receiver uses \(p(c_r\mid e_r)\propto p(c_r)g(e_{r,1}-c_{r,1})g(e_{r,2}-c_{r,2})\) over supported intervals. Endpoints are not clipped. Oracle intervals appear only in isolation studies. The target receiver conditions on noisy endpoints and sums intervals in \eqref{eq:pass_likelihood}.

All three receivers see identical \(D_{1:K}\). Let \(s\in\{\mathrm{sh},\mathrm{pass},\mathrm{obs}\}\) index the three scopes. The primary inferential unit is the within-replicate mean over five paired molecules because sites and molecules share calibration. For true label \(x_i^\star\), site APP \(p_i\), and projected APP \(\widetilde p_i\), the scores are
\begin{align}
\mathrm{NLL}&=-L^{-1}\sum_i\log p_i(x_i^\star),\nonumber\\
\mathrm{Brier}&=L^{-1}\sum_i\sum_x
 [p_i(x)-\mathbf1\{x=x_i^\star\}]^2,\nonumber\\
\mathrm{Acc}&=L^{-1}\sum_i
 \mathbf1\{\arg\max_xp_i(x)=x_i^\star\},\nonumber\\
\mathrm{TV}_{\max}&=\max_s L^{-1}\sum_i
 \tfrac12\|p_{s,i}-\widetilde p_{s,i}\|_1.
\label{eq:numerical_scores}
\end{align}
Expected NLL regret is \(L^{-1}\sum_i\KL(p_i\Vert\widetilde p_i)\). Contrasts are formed within replicates to preserve paired observations. At target scale, \(\Delta_b\) denotes pass-local minus shared cross-pass NLL and \(e_b=\Delta_b-\Delta_{\rm ref}\).

The 16-cell \(L=24,K=10\) grid crosses two settings each for \(q\), \(R\), events, and endpoints. Moderate and hard \((p_{\rm ins},p_{\rm del},p_{\rm dup})\) are \((0.03,0.05,0.08)\) and \((0.08,0.10,0.15)\). On offsets \(-2{:}2\), mild and moderate endpoint masses are \((0.025,0.10,0.75,0.10,0.025)\) and \((0.10,0.15,0.50,0.15,0.10)\). Span bins 6--10, 11--20, and 21--24 have masses \((0.40,0.30,0.30)\), followed by uniform span and start draws. The calibration design and length-four span law follow \eqref{eq:qhat}.

For each \((q,R)\), 30 replicates use 200 length-four, \(K=10\) calibration molecules to estimate \(q\) by \eqref{eq:qhat} and five paired targets per cell. The moderate-event, mild-endpoint estimate is shared across nuisance cells with the same \(q,R\). The three order-\(4\) scopes share observations and endpoint marginalization, use fitted \(\widehat q\), and marginalize latent intervals. Cross-pass NLL is primary, duplicate-local NLL is mechanistic, and Brier score and accuracy are descriptive.

Paired differences use two-sided 95\% Student-\(t\) intervals, with Holm adjustment for 16 primary cross-pass tests \cite{Holm_1979}. Duplicate-local intervals are unadjusted. A matched-law sensitivity at \(q=0.10\), uniform \(R\), moderate events, and moderate endpoints uses 100 acquisition-group bootstrap resamples. Each re-estimates \(E,q\) and generates 200 calibration molecules and five targets. We report the median, interquartile range, and exact 95\% Clopper--Pearson interval for the conditional positive-resample probability. All intervals are channel-conditional.

The crossed-law study fixes weighted \(R\) and pairs the two generating and decoder laws in Table~\ref{tab:emission} in two cells: \(q=0.10\) with moderate events and mild endpoints, and \(q=0.05\) with hard events and moderate endpoints. Each of four combinations has 30 paired replicates, 200 calibration molecules, and five \(L=24,K=10,b=4\) targets. Decoders share observations, fit \(q\) under the assumed \(E\), and form a four-test Holm family. Prespecified flags indicate boundary-fit rates above 5\%, opposite nonzero mean signs, or magnitude changes above \(0.005\) nats per residue or 20\% for matched magnitudes above \(0.01\).

\subsection{Order assessment and target-scale inference}
Order selection is separate from target comparison. In the \(L\le7\) order study, paired receivers use the same molecules and \(b=L-1\) provides the exact baseline. At \(L=24\), paired \(b=4,5\) receivers share observations and use the same \(\widehat q\), fitted on an independent calibration set.

For scope \(s\), let \(Z_s=U_{1:L}\) for shared inference and \(Z_s=X_{1:L}\) otherwise, with prior \(p_s\), exact pass likelihood \(\ell_{s,r}(z)=p_s(D_r\mid Z_s=z)\), and target \(\gamma_s(z)=p_s(z)\prod_r\ell_{s,r}(z)\). The normalized fused order-\(5\) chain \(Q_s=Q^s_{F,5}\) uses only observations and the fitted model. For \(A=7\) and \(q_{s,i}(a)=Q_s(Z_{s,i}=a)>0\), define
\begin{align}
Q_{s,i,a}(z)&=\frac{Q_s(z)\mathbf1\{z_i=a\}}{q_{s,i}(a)},\nonumber\\
M_s^{\rm LB}(z)&=\frac12Q_s(z)+\frac{1}{2LA}
\sum_{i=1}^{L}\sum_{a=1}^{A}Q_{s,i,a}(z).
\label{eq:label_blind_mix}
\end{align}
Balance weights \(w_m=\gamma_s(Z_s^{(m)})/M_s^{\rm LB}(Z_s^{(m)})\) are normalized as \(\bar w_m=w_m/\sum_jw_j\) to form sitewise APPs, with shared \(U\)-APPs mapped to \(X\) through \(\kappa\). This self-normalized deterministic-mixture estimator is LB--IS. Under \(M_s^{\rm LB}(z)>0\) whenever \(\gamma_s(z)>0\), it is consistent as all component sample sizes grow proportionally \cite{Hesterberg_1995,Elvira_2019}, although finite-sample bias remains. Proposal construction, target evaluation, and weighting are label-blind. Truth enters only scoring.

Per scope, baseline and site-symbol tilts total \(305{,}760+49(6{,}240)=611{,}520\) draws at \(L=7\) and \(1{,}048{,}320+168(6{,}240)=2{,}096{,}640\) at \(L=24\). Both use eight composition-balanced batches and likelihood batches of 512 sequences. LB--IS serves the fixed-condition unprojected-functional assessment, while the 16-cell study evaluates the practical finite-memory receiver.

The fixed high-allocation unprojected-model reference uses the same \(\gamma_s,Q_s\) and \(L\) truth-tail tilts. Let \(\eta_{s,i}(z_i)=\kappa(x_i^\star\mid z_i)\) for shared scope and \(\mathbf1\{z_i=x_i^\star\}\) otherwise, and set \(H_{s,i}=\mathbb E_{Q_s}\eta_{s,i}(Z_{s,i})\). Its mixture is \(M_s^{\rm TT}(z)=Q_s(z)[1/2+(2L)^{-1}\sum_i\eta_{s,i}(z_i)/H_{s,i}]\). Each scope uses \(1{,}048{,}320\) draws in eight batches. Truth enters the proposal but not the target functional. LB--IS and the reference use disjoint Monte Carlo streams on the same target realizations. The reference does not enumerate the full \(7^{24}\) posterior or provide a certified nonasymptotic error bound.

The target-scale comparison uses a primary condition (\(q=0.10\), weighted \(R\), moderate events, mild endpoints) and two diagnostics selected from the earlier order-\(4\) study: its smallest effect (\(q=0.05\), weighted \(R\), hard events, moderate endpoints) and largest shared NLL (\(q=0.10\), uniform \(R\), hard events, moderate endpoints). The reference construction, conditions, allocations, criteria, and reported quantities were fixed before LB--IS evaluation. Each condition comprises ten paired target realizations with separate calibration sets. Target realizations are the inferential units, with sites treated as within-realization observations. The paired design compares estimators on common targets using disjoint Monte Carlo streams.

For weighted draws, \(N_{\rm eff}=(\sum_m w_m)^2/\sum_m w_m^2\). Precision requires \(N_{\rm eff}\ge10^4\), \(\max_m\bar w_m\le0.005\), maximum APP standard error \(\le0.005\), split-half mean/maximum APP-TV \(\le0.01/0.02\), and cross-pass Monte Carlo standard error/split-half difference \(\le0.0025/0.01\) nats per residue. Relative errors also require \(|\Delta_{\rm ref}|>0.01\) and split-half difference \(\le0.25|\widehat\Delta|\). A sign is reportable when its eight-batch \(t_7\) interval excludes zero with half-width \(\le\min\{0.005,0.20|\widehat\Delta|\}\).

Within each replicate and score, observation-local minus shared equals cross-pass plus duplicate-local. Calibration and target molecules are disjoint, independent draws evaluate calibration bias and root-mean-square error (RMSE), and inferential intervals use replicate summaries.

\subsection{Approximation and calibration assessment}
In Table~\ref{tab:target_reference}, agreement requires absolute cross-pass NLL-contrast error \(\le0.005\) nats per residue, relative error \(\le20\%\) for \(|\Delta_{\rm ref}|>0.01\), mean/90th-percentile maximum branch-APP TV \(\le0.005/0.02\), expected NLL regret \(<0.001\), and no sign reversal above the effect floor. Targets are exact at \(L=7\) and use the high-allocation reference at \(L=24\). Sitewise TV and regret are reduced to scope-realization maxima before ten-realization one-sided bounds. The joint agreement label is Within when every criterion is met, Inconclusive when precision is insufficient or an eligible effect remains unresolved, and Outside when any resolved criterion exceeds its prespecified tolerance. Orders above \(L-1\) are truncated.

\begin{figure*}[!t]
\centering
\resultfigure{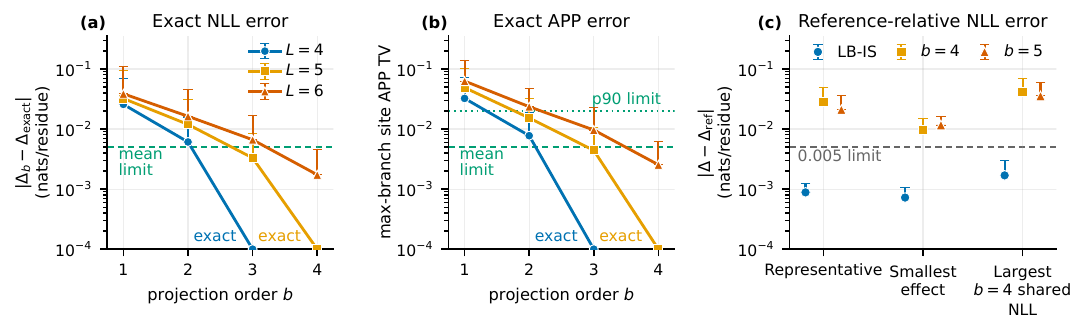}
\caption{Order assessment. (a), (b) Exact \(K=10\) means and empirical 90th-percentile upper whiskers for \(L=4,5,6\). Here \(\Delta_{\rm exact}\) denotes the \(b=L-1\) cross-pass contrast. (c) Absolute cross-pass NLL errors relative to the fixed high-allocation unprojected-model reference for LB--IS and orders 4 and 5 in three selected \(L=24,K=10\) conditions. Aggregate whiskers are one-sided 95\% Student-\(t\) upper bounds over ten realizations. The line marks the \(0.005\)-nat-per-residue criterion.}
\label{fig:order_evidence}
\end{figure*}

Exact endpoint-marginalized enumeration covers \(L=4,5,6\), all three scopes, nested \(K\), and every order through the exact \(b=L-1\) baseline. Fig.~\ref{fig:order_evidence}(a)--(b) summarizes the most demanding \(K=10\) results. With \(b=\min\{b_{\max},L-1\}\), the exact strata select \(b_{\max}=4\) as the smallest tested common cap satisfying every stated per-stratum criterion over \(L=4,5,6\), with no eligible sign reversal. Order \(b=2\) misses several criteria, while \(b=3\) misses the \(L=6\), \(K\in\{5,10\}\) strata. Order 4 is exact at \(L=4,5\) and genuinely approximate at \(L=6\). At \(L=7\), order 4 retains contrast and regret agreement, with mean APP-TV slightly above its prespecified tolerance.

The label-blind \(L=7\) benchmark contains nine fixed cases, three per condition, and all 27 scope estimates. Exact \(7^7\) enumeration supplies the posterior, and an independent dynamic-programming formulation evaluates the target. Because scaled \(L=7\) spans extend beyond the \(L=24\) proposal construction, \(Q_s\) uses exact per-pass contiguous moments. All precision, posterior-agreement, and score-agreement criteria are satisfied. Maximum mean, 90th-percentile, and sitewise APP-TV are \(0.000779\), \(0.001268\), and \(0.002107\), maximum expected-NLL regret is \(2.84\times10^{-6}\), and maximum case cross-pass error is \(0.002522\) nats per residue. These exact-scale results validate sampling, reweighting, and back-mapping on the fixed \(L=7\) benchmark. Target-scale behavior is assessed separately against the fixed high-allocation \(L=24\) numerical reference.

\begin{table*}[!t]
\caption{Agreement with the fixed high-allocation self-normalized Monte Carlo reference for the unprojected model at \(L=24,K=10\). Entries are means [one-sided 95\% Student-\(t\) upper bounds] over ten realizations. Relative-error bounds use \(t_{n_r-1}\) over the \(n_r\) reportable cases. All other aggregate bounds use \(t_9\). APP-TV and regret are scope maxima. The dagger denotes a descriptive relative-error summary over 6 of 7 eligible effects. Joint reports Within, Outside, or Inconclusive under the complete criterion set. The last condition maximizes order-4 shared NLL. NLL is per residue.}
\label{tab:target_reference}
\centering
\footnotesize
\setlength{\tabcolsep}{0.8pt}
\renewcommand{\arraystretch}{0.98}
\begin{tabular*}{\textwidth}{@{\extracolsep{\fill}}lcccccccc}
\toprule
Condition & Estimator & \shortstack{Abs. contrast\\error [U]} & \shortstack{Relative\\error [U]} & \shortstack{Mean APP-TV\\{}[U]} & \shortstack{90th-pct. APP-TV\\{}[U]} & \shortstack{NLL regret\\\((\times10^{-3})\) [U]} & \shortstack{Relative/sign\\status} & Joint \\
\midrule
Representative & LB--IS & \(.000881\,[.001240]\) & \(.00355\,[.00499]\) & \(.000544\,[.000673]\) & \(.001199\,[.001391]\) & \(.00563\,[.00690]\) & Pass & Within \\
 & Ord. 4 & \(.02805\,[.04943]\) & \(.09140\,[.14394]\) & \(.02913\,[.03789]\) & \(.09209\,[.11823]\) & \(8.28\,[12.05]\) & Pass & Outside \\
 & Ord. 5 & \(.02102\,[.03571]\) & \(.07057\,[.10500]\) & \(.02093\,[.02969]\) & \(.06107\,[.08946]\) & \(5.25\,[8.93]\) & Pass & Outside \\
\addlinespace[1pt]
Smallest effect & LB--IS & \(.000723\,[.001079]\) & \(.01774\,[.03709]^{\dagger}\) & \(.001267\,[.001483]\) & \(.002022\,[.002341]\) & \(.0148\,[.0187]\) & Inconclusive & Inconclusive \\
 & Ord. 4 & \(.00966\,[.01490]\) & \(\mathrm{Inconclusive}\) & \(.08494\,[.09674]\) & \(.17566\,[.22110]\) & \(36.24\,[46.94]\) & \(\mathrm{Inconclusive}\) & Outside \\
 & Ord. 5 & \(.01168\,[.01620]\) & \(\mathrm{Inconclusive}\) & \(.06506\,[.07740]\) & \(.14879\,[.19740]\) & \(22.96\,[32.56]\) & \(\mathrm{Inconclusive}\) & Outside \\
\addlinespace[1pt]
\shortstack[l]{Largest \(b=4\)\\shared NLL} & LB--IS & \(.001692\,[.003040]\) & \(.01276\,[.02174]\) & \(.001232\,[.001479]\) & \(.002115\,[.002444]\) & \(.0130\,[.0167]\) & Pass & Within \\
 & Ord. 4 & \(.04166\,[.06891]\) & \(.22442\,[.31577]\) & \(.08052\,[.09591]\) & \(.15946\,[.19093]\) & \(32.91\,[42.56]\) & Fail & Outside \\
 & Ord. 5 & \(.03556\,[.05922]\) & \(.21930\,[.29865]\) & \(.05782\,[.07515]\) & \(.12787\,[.17238]\) & \(20.03\,[30.28]\) & Fail & Outside \\
\bottomrule
\end{tabular*}
\end{table*}

Every aggregate absolute marginal-posterior and score-agreement criterion is met in all three conditions. Joint LB--IS agreement is Within for the representative and largest-order-4-shared-NLL conditions, and Inconclusive for the smallest effect because its relative error and sign are unresolved. In the smallest-effect condition, relative error is reportable for six of seven eligible effects. Five signs are reportable, and all five agree with the reference. One high-NLL realization has absolute contrast error \(0.00809\), so the agreement classification is aggregate rather than realization-uniform. Across the six estimator-condition comparisons, the finite-order mean absolute contrast errors are 13.4--31.8 times the LB--IS errors relative to the same numerical reference.

Together, these results separate the exact short-sequence regime from target-scale receiver comparisons. Order 4 is the smallest common cap satisfying all exact-enumeration \(L\le6\) criteria. At \(L=24\), order 5 reduces several errors, but neither finite order attains joint reference agreement in the three conditions. The order-5-minus-order-4 NLL changes are \(0.00393\mathbin{\pm}0.00357\), \(-0.00127\mathbin{\pm}0.00182\), and \(0.00691\mathbin{\pm}0.00511\) nats per residue. Within this selected family, the representative and largest-shared-NLL contrasts are resolved, with nominal Holm-adjusted \(p\)-values of \(0.0084\) and \(0.0296\), while the smallest-effect contrast is unresolved (\(p=0.8172\)). The 16-cell order-4 study therefore characterizes the specified finite-memory receiver.

\begin{figure}[!b]
\centering
\includegraphics[width=\columnwidth]{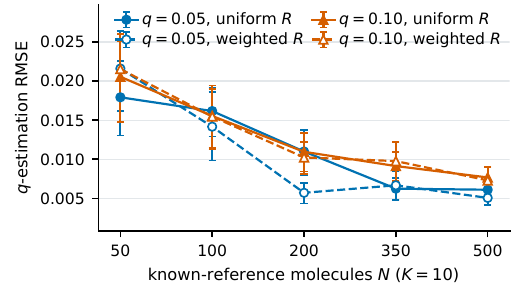}
\caption{Fixed-family calibration with exact likelihood evaluation. RMSE uses 30 independently simulated reference sets. Bars are 95\% percentile-bootstrap intervals from 10,000 resamples of set-level errors. The PASTOR-derived \(E\) is fixed while \(q,R\) vary. Lines connect the five evaluated \(N\) values.}
\label{fig:calibration_scaling}
\end{figure}

The receiver-study calibration ensemble yields RMSE \(0.00890\)--\(0.01146\) at \(N=200,K=10\), with no boundary fits and coefficient evaluation agreeing with direct enumeration within \(1.4\times10^{-14}\). In the separate scaling ensemble, RMSE across the four \((q,R)\) cells in Fig.~\ref{fig:calibration_scaling} decreases from \(0.01792\)--\(0.02160\) at \(N=50\) to \(0.00506\)--\(0.00767\) at \(N=500\). At fixed \(N=200\), increasing \(K\) from 2 to 10 reduces RMSE from \(0.01666\)--\(0.02418\) to \(0.00571\)--\(0.01099\). All 36 cells avoid the boundary. These matched-model simulations establish fixed-family self-consistency. Molecule-linked observations would support empirical persistence calibration within the specified family. The identifiability diagnostics give \(\|\Delta_R\|_F=0.0284\)--\(0.0385\), Fisher information \(0.0195\)--\(0.0234\), and two-pass TV \(0.587\)--\(0.623\).

\begin{figure*}[!t]
\centering
\resultfigure{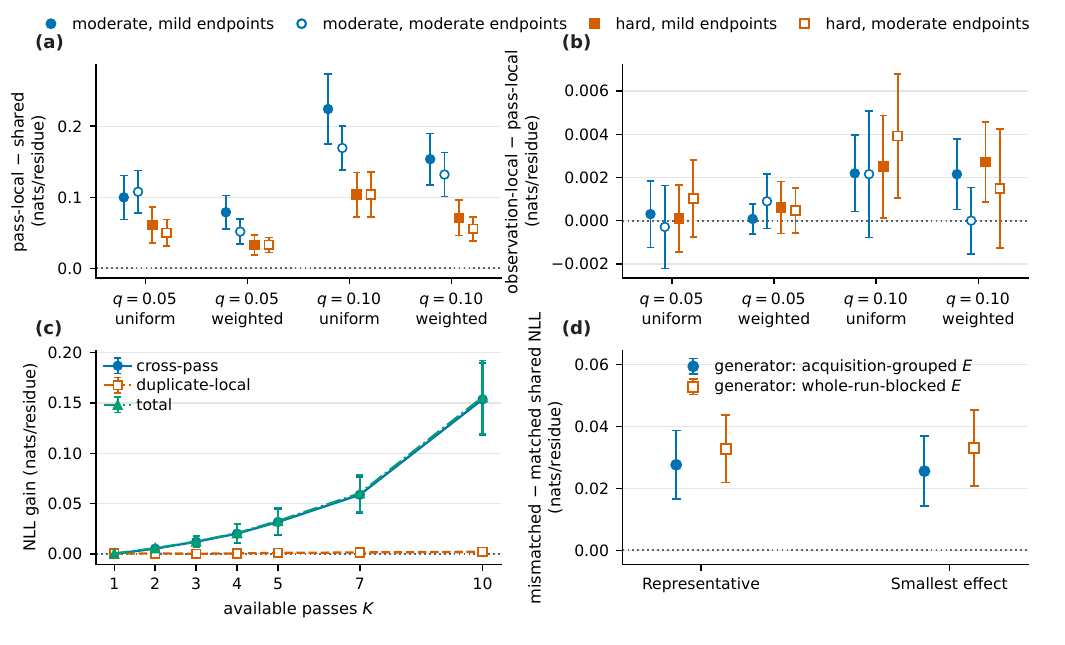}
\caption{Order-\(4\) receiver results. (a), (b) Cross-pass and duplicate-local NLL at \(L=24,K=10\). (c) Nested-\(K\) decomposition over common molecules and first-\(K\) pass prefixes. (d) Four prespecified crossed-law emission-mismatch contrasts. Positive values favor shared decoding in (a), pass-local decoding in (b), and matched decoding in (d). Lines connect the seven evaluated \(K\) values. Bars are unadjusted paired two-sided 95\% Student-\(t\) intervals over 30 replicates. Holm-adjusted results appear in the text. All differences include branch-specific projection error.}
\label{fig:deployment}
\end{figure*}

\subsection{Target-scale finite-memory receiver comparisons}
Fig.~\ref{fig:deployment} shows that, with the specified order-\(4\) receiver, the shared branch has \(0.033\)--\(0.224\) nats per residue lower NLL than the pass-local branch across 16 cells after Holm adjustment. In the representative nested-prefix series, the cross-pass contrast rises from zero at \(K=1\) to \(0.03175\) at \(K=5\) and \(0.15367\) at \(K=10\). Duplicate-local differences span \(-0.0003\)--\(0.0039\), with five unadjusted intervals favoring pass-local decoding. Their sum is the total contrast. These receiver-level intervals include branch-specific projection error. Cross-pass Brier favors shared in 11 cells, with no resolved branch advantage in five. Duplicate-local Brier never favors pass-local dependence. Accuracy favors shared inference in one cell and is unresolved in fifteen.

Matched-law resampling gives median cross-pass NLL \(0.1689\), interquartile range \([0.0861,0.2484]\), and 99 positive values among 100, with exact conditional-probability interval \([0.9455,0.9997]\). Median \(q\)-error is \(-0.0006\), with interquartile range \([-0.0075,0.0078]\). Emission mismatch raises shared NLL by \(0.0255\)--\(0.0331\) in all four Holm-significant comparisons. Every cross-pass magnitude changes materially without boundary instability or sign reversal. These results quantify sensitivity to the evaluated emission pair.

\section{Discussion and Limitations}
In the aligned specialization, independent redraw may preserve the MAP action while true-label NLL grows with \(K\). This matters for posterior-magnitude decisions such as threshold stopping and database matching. Their application-specific losses remain outside the present evaluation.

Exact \(L\le6\) benchmarks select order 4 as the smallest common tested cap, and exact \(L=7\) enumeration validates LB--IS. At \(L=24\), LB--IS meets all aggregate absolute marginal-posterior and score criteria and resolves representative and high-NLL contrasts. The reference supports selected unprojected functionals, whereas neither order 4 nor 5 meets every joint criterion. Thus, the order-4 grid remains a finite-memory receiver study.

Five runs per order on one fixed \(L=24,K=10\) input yielded median decoder times of 13.8/91.2 s and median peak-memory increments of 57.2/150.6 MiB for orders 4/5. These machine-dependent measurements characterize burden, not throughput. Calibration identifies \(q\) for fixed \(E,R\), and the crossed-law study covers two emission laws. PASTOR informs the seven-residue law and motivates partial rereading, while persistence is modeled. Molecule-linked splits with retained nulls, state-dependent coverage checks, and proper scores define the next empirical validation step. Extensions include pass-evolving states, electrical correlation, structured priors, and context-dependent emissions.

\section{Conclusion}
Compound-pass inference gives the equivalence-class limit, while aligned observation-local redraw yields linear-in-\(K\) true-label NLL growth. Exact \(L=7\) enumeration validates LB--IS on the fixed benchmark. At \(L=24,K=10\), representative and high-NLL contrasts resolve, but the near-zero contrast remains inconclusive. Order 4 is the smallest tested cap meeting all exact \(L\le6\) criteria, but orders 4 and 5 do not attain joint target-scale agreement. Across 16 cells, order-4 shared decoding lowers NLL by \(0.033\)--\(0.224\) nats per residue relative to pass-local. Molecule-linked data would enable empirical persistence calibration.

\appendices
\section{Proofs}
\label{app:proofs}
\paragraph{Proof of Theorem~\ref{thm:shared}}
Fix \(u\in\cU_+\). Any \(v\ne u\) whose support excludes an \(E_u\)-positive outcome is eventually eliminated, while every other incorrect state's covered-sample log-likelihood ratio converges almost surely to \(-\KL(E_u\Vert E_v)<0\). Finiteness and \(N_K\to\infty\) concentrate the posterior on \(u\), or on its coincident-row class. The Markov relation, continuity of the maximum and entropy, and bounded convergence give \eqref{eq:error_floor} and \eqref{eq:shared_nll_limit}.

\paragraph{Proof of Theorem~\ref{thm:local}}
Conditional on \(U=u\), the strong law gives \(K^{-1}\sum_{k:B_k=1}\log M_x(Y_k)\to-h[H(E_u)+d_x(u)]\). Denominator exponents yield concentration on \(\mathcal S(u)\) and slope \(h[d_x(u)-d_\star(u)]\). Finiteness and the positive decoder prior give \(m_0=\min_{x,a:M_x(a)>0}M_x(a)>0\) and minimum prior mass \(\alpha_0>0\). The support assumption yields \(0\le-K^{-1}\log p_{\rm obs}(X\mid O_{1:K})\le-K^{-1}\log\alpha_0-\log m_0\). Dominated convergence proves \eqref{eq:local_nll_slope}.

\paragraph{Proof of Theorem~\ref{thm:coverage}}
For fixed \(u\), excess action risk is zero when \(a(v)=a(u)\). For every \(v\) inducing another action, Markov's inequality bounds its selection probability by \((p_U(v)/p_U(u))^t[\sum_o\widetilde E_u(o)^{1-t}\widetilde E_v(o)^t]^K\). A finite union bound and optimization over \(t\in(0,1)\) prove \eqref{eq:achievable_exponent}. Prior ratios have zero exponent. Observing \(U\) lower-bounds risk, and direct Bayes decoding is no worse than the two-stage rule.

\paragraph{Proof of Proposition~\ref{prop:compound}}
Any \(c'\ne c\) whose support excludes a \(W_c\)-positive outcome is eventually eliminated, while every other incorrect class has normalized log-likelihood ratio converging to \(-\KL(W_c\Vert W_{c'})<0\). Finiteness concentrates the posterior on \(c\), and equal within-class likelihoods preserve conditional prior ratios. Marginalization, continuity of the maximum and entropy, and bounded convergence prove \eqref{eq:compound_posterior}--\eqref{eq:compound_nll}.

\paragraph{Proof of Corollary~\ref{cor:compound_exponent}}
Given \(V=c\), all state sequences in the class have law \(W_c\), hence \(X_{1:L}\perp D_{1:K}\mid V\). Replacing \((U,E_u,a(u))\) by \((V,W_c,a(c))\) in Theorem~\ref{thm:coverage}'s argument gives \eqref{eq:compound_achievable}. The risk ordering follows from the informativeness of \(V\) and optimality of direct Bayes decoding.

\paragraph{Proof of Theorem~\ref{thm:order}}
Maximizing \(\mathbb E_P\log Q\) selects the order-\(b\) prefix and positive-context conditionals of \(P\). The chain rule gives \eqref{eq:cmi_residual}, monotonicity, and \(Q_{L-1}=P\). With \(r_r(u)=\log[P_r(u)/Q_{r,b}(u)]\), prior cancellation proves \eqref{eq:odds_bound}. For \(\rho=P_F/Q_{F,b}\), let \(a=\min\rho\), \(c=\max\rho\). Then \(a\le1\le c\), \(c/a\le e^{\Omega_b}\), and convexity gives \(\operatorname{TV}(P_F,Q_{F,b})\le(c-1)(1-a)/(c-a)\le\tanh(\Omega_b/4)\). Equality in the last maximization occurs at \(a=e^{-\Omega_b/2},c=e^{\Omega_b/2}\). Markov kernels contract TV. Support mismatch leaves the trivial bound.

\paragraph{Proof of Proposition~\ref{prop:pair}}
Marginalization and \eqref{eq:transition} yield \(G_q=G_0+q\Delta_R\), proving \eqref{eq:pair_affine} and identifiability exactly when \(\Delta_R\ne0\). Included pairs have \(\mathbb E\widehat G=G_q\), so \eqref{eq:pair_estimator} is unbiased before bounded projection.

\renewcommand{\IEEEbibitemsep}{-0.5pt}
\bibliographystyle{IEEEtran}
\bibliography{references}
\end{document}